\documentclass[submission, copyright, creativecommons]{eptcs}
\providecommand{\event}{CAPNS 2018} 

\usepackage{amsmath}
\usepackage{amsfonts}

\usepackage{amsthm}
\newtheorem{definition}{Definition}

\usepackage{bussproofs} 

\usepackage{color}

\usepackage{tikz}
\usetikzlibrary{shapes.geometric, decorations.markings}
\tikzset{->-/.style={decoration={markings, mark=at position .5 with {\arrow{>}}}, postaction={decorate}}}

\newcommand\OG{\mathbf{OG}} 
\newcommand\FVect{\mathbf{FVect}}
\newcommand\Rel{\mathbf{Rel}}

\newcommand\semantics[1]{[\![#1]\!]}
\newcommand\define[1]{\textit{#1}}

\title{Towards Functorial Language-Games}
\author{Jules Hedges \institute{Quantum Group, University of Oxford} \email{julian.hedges@cs.ox.ac.uk} \and Martha Lewis \institute{ILLC, University of Amsterdam} \email{m.a.f.lewis@uva.nl}}

\def\titlerunning{Towards Functorial Language-Games}
\def\authorrunning{J. Hedges \& M. Lewis}

\begin{document}
\maketitle

\begin{abstract}
	In categorical compositional semantics of natural language one studies functors from a category of grammatical derivations (such as a Lambek pregroup) to a semantic category (such as real vector spaces). We compositionally build game-theoretic semantics of sentences by taking the semantic category to be the category whose morphisms are open games. This requires some modifications to the grammar category to compensate for the failure of open games to form a compact closed category. We illustrate the theory using simple examples of Wittgenstein's language-games.
\end{abstract}

\section{Introduction}
A key component of language is that it is used by agents to communicate about the world they live in, and that it forms part of a network of actions used to interact with the world. This notion was used by Wittgenstein in his description of language games as a way of exploring how signs and symbols could come to refer to objects in the world \cite{wittgenstein_philosophical_investigations}. It has also been used in the artificial intelligence community as a way of enabling artificial systems to develop their own meanings and to examine how those meanings arise \cite{steels2003}. Relatedly, game theory has been applied to language research. \cite{jager2008} provides a review of how game theory has been  to formalise the idea of communication as a signalling game, to examine the notion of relevance, and, using evolutionary game theory, to investigate how signals evolve to correspond to a convex region of representational space.

Another approach to language is the field of computational linguistics, and in particular compositional distributional semantics such as that presented in \cite{coecke2010, paperno2014}. This approach seeks to unify the rich semantic representations that can be learnt from text corpora with the compositional and generative power of formal semantics. However, it is agent-free. There is no sense of language being used by an agent, of its relation to the world, or of its part in a network of actions. To use ideas from compositional distributional semantics in a system interacting with the world, the theory must be developed further.

In this paper we begin to unify the categorical compositional distributional semantics of \cite{coecke2010} with the categorical approach to game theory introduced in \cite{hedges_etal_compositional_game_theory, hedges_towards_compositional_game_theory}. We use the category of open games \cite{hedges_etal_compositional_game_theory} to represent the semantics of words and phrases. However, the grammar used in \cite{coecke2010} is not available to used, since the category of open games is not compact closed. We therefore develop a novel grammar, similar to Lambek's protogroup grammar \cite{lambek1997}, and which is based on the free teleological category introduced in \cite{hedges_coherence_lenses_open_games}. Finally, we give an example based on one of Wittgenstein's language games.

\section{Background}
\subsection{Categorical Compositional Semantics}\label{sec:DisCoCat}
In the study of language at least two distinct camps can be identified. On one side, the way that words compose to form phrases and sentences is the object of research. On the other, the meanings of the words themselves is seen as paramount. These two approaches can be labelled syntax and semantics respectively. There is then also pragmatics, the aspect of how words, phrases and so on are actually used in discourse, but we do not deal with this at present. The categorical compositional approach was introduced in the context of distributional semantics by \cite{coecke2010}, and further elucidated in \cite{bolt2017}. The approach was outlined in \cite{bolt2017} as follows:
\begin{enumerate}  
  \label{item:compstruct}
\item \begin{enumerate}
\item Choose a compositional structure, such as a pregroup, or other categorial grammar.
\item Interpret this structure as a category, the \define{grammar category}.
\end{enumerate}
\item \begin{enumerate}
  \label{item:meaningspace}
\item Choose or craft appropriate meaning or concept spaces, such as vector spaces.
  \label{item:meaningcategory}
\item Organize these spaces into a category, the \define{semantics category}, with the same abstract structure as the grammar category.
\end{enumerate}
  \label{item:interpret}
\item Interpret the compositional structure of the grammar category in the semantics category via a functor preserving the type reduction
  structure.
  \label{item:reduction}
\item Bingo! This functor maps type reductions in the grammar category onto algorithms for composing meanings in the semantics category. 
\end{enumerate}
In \cite{coecke2010}, the authors instantiated this approach in the context of compact closed categories using pregroup grammar as the grammar category and $\FVect$ or $\Rel$ as the semantic category. \cite{coecke2013} show how the grammar category may be changed to Lambek categorial grammar. In \cite{piedeleu2015, balkir2015}, the same approach was instantiated using pregroup grammar and the category of completely positive maps on $\FVect$, and \cite{bolt2017} introduce a category of convex algebras and convex relations as the semantics category. 

The key concept in all of the above examples is that of a \emph{monoidal category}. This is a category with a parallel composition operation $\otimes$, which is associative, has a unit, and acts bifunctorially on objects and morphisms. Monoidal categories have a convenient diagrammatic calculus, which we will make use of in later sections. 
For details of this calculus, see for example \cite{coeckepaquette}.

Many of the examples work in the framework of compact closed categories. A compact closed category is a monoidal category with left and right adjoints $(-)^l$, $(-)^r$ on objects and the morphisms:
\[
\label{eq:unitsandcounits}
\epsilon^l_A:A^l \otimes A \rightarrow I_A, \qquad \epsilon^r_A: A \otimes A^r \rightarrow I_A, \qquad \eta_A^l: I_A \rightarrow A \otimes A^l, \qquad \eta_A^r : I_A \rightarrow A^r \otimes A
\]
termed counits and units respectively, which satisfy the \emph{snake equations}
\begin{align*}
&(1_A \otimes \epsilon^l )\circ (\eta^l \otimes 1_A) = 1_A &\:& (\epsilon^r \otimes 1_A) \circ (1_A \otimes \eta^r ) = 1_A
\\
&(\epsilon^l \otimes 1_{A^l})\circ (1_{A^l} \otimes \eta^l)= 1_{A^l} &\:& (1_{A^r} \otimes \epsilon^r ) \circ (\eta^r \otimes 1_{A^r} ) = 1_{A^r}
\end{align*}
In compact closed categories, the diagrammatic calculus is enhanced by the fact that we can bend wires.

In our linguistic examples, we frequently use the notion of a pregroup grammar. A pregroup grammar is built from the free compact closed category over a set of basic types. Consider the set $\{n, s\}$ of types corresponding to nouns and declarative sentences. The tensor product is denoted by concatenation, allowing us to build composite types. For example, the type  of a transitive verb is $n^r s n^l$. A string of types is judged \emph{grammatical} if it reduces to the type $s$ under application of the unit and counit morphisms. So, the sentence $\textit{Gamblers take chances}$ is typed: $n \;n^rsn^l\;n$ and reduces as follows:
\[ n \;n^rsn^l\;n \xrightarrow{\epsilon^r_n \; 1_s\;\epsilon^l_n} sn^l\;n \xrightarrow{1_s\;\epsilon^l_n} s \]

More precisely, a pregroup grammar is a triple $G = (\mathcal L, T, s, \mathcal D)$ where:
\begin{itemize}
	\item $\mathcal L$ is the \emph{lexicon}, a set of words
	\item $T$ is a set of \emph{basic types}
	\item $s \in T$ is the distinguished \emph{sentence type}
	\item $\mathcal D \subseteq \mathcal L \times P_T$ is the \emph{dictionary}, where $P_T$ is (the set of elements of) the free pregroup on $T$.
\end{itemize}
We model the grammar in the category $\mathbf{FVect}$ of finite dimensional vector spaces. We must choose a finite-dimensional vector space $\semantics t$ for every $t \in T$.
This induces a monoidal functor $\semantics - : \mathcal C_T \to \mathbf{FVect}$, where $\mathcal C_T$ is the free compact closed category with generating objects $T$, by specifying that $\semantics{t^l} = \semantics{t^r} = \semantics t^*$ (where $-^*$ denotes the dual vector space), and $\semantics{t_1 t_2} = \semantics{t_1} \otimes \semantics{t_2}$.
(Note the careful distinction between $P_T$ and $\mathcal C_T$, to avoid the argument in \cite{preller_logical_distributional_models} that monoidal functors out of $P_T$ are necessarily trivial.)
We must also specify a word-vector $\semantics w \in \semantics t$ for each word $w \in \mathcal L$ with $(w, t) \in \mathcal D$; in practice these vectors will be computed from corpora of text by statistical or machine learning methods.
Now given a sentence $w_1 \ldots w_n$, we compute its semantic vector $\semantics{w_1 \ldots w_n} \in \semantics s$ by the following steps:
\begin{enumerate}
	\item Choose a type $(w_i, t_i) \in \mathcal D$ for each word (this is automatic if each word in $\mathcal L$ is assigned a unique type by $\mathcal D$)
	\item Choose a morphism $f : t_1 \cdots t_n \to s$ in $\mathcal C_T$, which represents a grammatical parsing of the sentence
	\item Apply the linear map $\semantics f : \semantics{t_1} \otimes \cdots \otimes \semantics{t_n} \to \semantics s$ to the bag-of-words vector $\semantics{w_1} \otimes \cdots \otimes \semantics{w_n}$
\end{enumerate}

In the current paper, we instantiate the approach with the category $\OG$ of open games as the semantics category rather than $\mathbf{FVect}$. This allows us to express commands and subsequent actions, and is therefore a richer setting than previously, in which there was no concept of agent, nor of action. The category of open games, however, is not compact closed. In particular it has counits, but no units. We therefore use the notion of a \emph{protogroup} \cite{lambek1997} as the basis for our grammar category.

\subsection{Language Games}\label{sec:language-games}
The approach to language outlined in \cite{coecke2010} is focussed on solving a particular problem in computational linguistics. However, language in general is used by agents, in the world, and together with other actions. This attitude to language was outlined in \cite{wittgenstein_philosophical_investigations}. Meanings of words arise from the situations in which they are used. The term \define{language game} is used to talk about small fragments of language that can be used to highlight how language obtains its meaning. 

This approach to language has been used as a way of bootstrapping language learning in artificial intelligence approaches. These are summarised in \cite{steels2003}. In brief, rather than attempt to specify representations of concepts for AI systems, these systems are given the ability to form their own representations, based on their interactions with the world and with other systems (including humans). The Talking Heads experiment \cite{steels1998origins} was designed to show how a population of interacting agents could develop a shared lexicon, together with perceptually grounded categorizations corresponding to items of the lexicon. The game is played as follows. Both the speaker and the hearer can see the same set of shapes. The speaker chooses a shape, and says some words corresponding to that shape. Then, the hearer must point out which shape it guesses. If the hearer gets the shape wrong, the speaker indicates which shape is the correct one. This game is played a number of times, with each agent playing both the role of the speaker and the hearer in different rounds. As the game is played, representations are built up and altered, and the sets of words and concepts of the agents become coordinated, without central control or telepathy.

The language game we will outline in the current paper is based on Wittgenstein's builder and assistant game. This will be given in more detail in section \ref{sec:example}. Briefly, the game runs as follows:
\begin{quote}
The language is meant to serve for communication between a builder A and an assistant B. A is building with building-stones: there are blocks, pillars, slabs and beams. B has to pass the stones, in the order in which A needs them. For this purpose they use a language consisting of the words ``block", ``pillar", ``slab", ``beam". A calls them out; — B brings the stone which he has learnt to bring at such-and-such a call. Conceive this as a complete primitive language. \cite[\S 2]{wittgenstein_philosophical_investigations}
\end{quote}

In contrast to the Talking Heads game, the game here is not played iteratively. The assistant $B$ has already learnt what each word means. We write down how the language game can be formalized game-theoretically, and give an extension of it, in which imperative verbs are introduced.

%


%

\subsection{Open games}

A game, informally speaking, is a model of (any number of) interacting agents, in which each individual agent acts \emph{rationally} or \emph{strategically} in order to optimise some outcome that is personal to them.
The notion of \emph{Nash equilibrium} is used to describe an assignment of behaviours to each agent which are mutually rational, that is, each is rational under the assumption that the others are fixed.

An open game is a piece of a game, which can be composed together to build ordinary games \cite{hedges_towards_compositional_game_theory,hedges_etal_compositional_game_theory}.
There is a symmetric monoidal category $\mathbf{OG}$ whose morphisms are (equivalence classes of) open games, in which the categorical composition and monoidal product are respectively sequential and simultaneous play of open games.

The objects of the category $\mathbf{OG}$ are pairs of sets $\binom X S$, which represent sets of ordinary forward-flowing information, and `counterfactual' information that appears to flow backwards.
\begin{definition}
	Let $X, S, Y, R$ be sets.
	An \emph{open game} $\mathcal G : \binom X S \to \binom Y R$ is a 4-tuple $(\Sigma_\mathcal G, \mathbf P_\mathcal G, \mathbf C_\mathcal G, \mathbf E_\mathcal G)$ where:
	\begin{itemize}
		\item $\Sigma_\mathcal G$ is a set, the set of \emph{strategy profiles} (roughly, possible behaviours of $\mathcal G$ unconstrained by rationality)
		\item $\mathbf P_\mathcal G : \Sigma_\mathcal G \times X \to Y$ is the \emph{play function}, which runs a strategy profile on an observation to produce a choice
		\item $\mathbf C_\mathcal G : \Sigma_\mathcal G \times X \times R \to S$ is the \emph{coplay function}, which propagates payoffs further backwards in time
		\item $\mathbf E_\mathcal G : X \times (Y \to R) \to \mathcal P (\Sigma_\mathcal G)$ is the \emph{equilibrium function}, which gives a subset of $\mathbf E_\mathcal G (h, k) \subseteq \Sigma_\mathcal G$ of strategy profiles that are Nash equilibria for each \emph{context} $(h, k)$. The context consists of a \emph{history} $h : X$ (which says what happened in the past) and a \emph{continuation} $k : Y \to R$ (which says what will happen in the future given the present).
	\end{itemize}
\end{definition}

The string diagram language for $\mathbf{OG}$ has directed strings, where a pair $\binom X S$ is represented by an $X$-labelled string directed forwards together with an $S$-labelled string directed backwards.
A general open game $\mathcal G : \binom X S \to \binom Y R$ is represented by a string diagram of the form
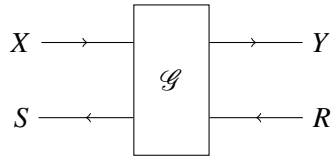
\begin{center} \begin{tikzpicture}
	\node (X) at (0, .5) {$X$}; \node (S) at (0, -.5) {$S$}; \node (Y) at (4, .5) {$Y$}; \node (R) at (4, -.5) {$R$};
	\node (G) [rectangle, minimum height=2cm, minimum width=1cm, draw] at (2, 0) {$\mathcal G$};
	\draw [->-] (X) to (G.west |- X); \draw [->-] (G.east |- Y) to (Y); \draw [->-] (R) to (G.east |- R); \draw [->-] (G.west |- S) to (S);
\end{tikzpicture} \end{center}

A \emph{closed game} is a scalar $\mathcal G : I \to I$ in $\mathbf{OG}$, where $I = \binom 1 1$ is the monoidal unit.
A closed game $\mathcal G$ is determined by a set $\Sigma_\mathcal G$ of strategy profiles and a subset $\mathbf E_\mathcal G \subseteq \Sigma_\mathcal G$ of Nash equilibria.

As an example, consider a single decision made by an agent who observes an element of $X$ before choosing an element of $Y$, in order to maximise a real number.
This situation is represented by the open game $\mathcal A : (X, 1) \to (Y, \mathbb R)$ defined by the following data:
\begin{itemize}
	\item The set of strategy profiles is $\Sigma_\mathcal A = X \to Y$
	\item The play function is $\mathbf P_\mathcal A (\sigma, x) = \sigma (x)$
	\item The coplay function has codomain $1$, hence is trivial
	\item The equilibrium function is $\mathbf E_\mathcal A (h, k) = \{ \sigma : X \to Y \mid \sigma (h) \in \arg\max k \}$
\end{itemize}
We represent $\mathcal A$ by the string diagram
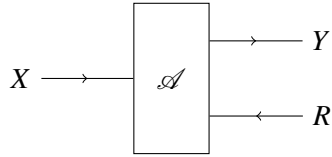
\begin{center} \begin{tikzpicture}
	\node (X) at (0, 0) {$X$}; \node (Y) at (4, .5) {$Y$}; \node (R) at (4, -.5) {$R$};
	\node (G) [rectangle, minimum height=2cm, minimum width=1cm, draw] at (2, 0) {$\mathcal A$};
	\draw [->-] (X) to (G.west |- X); \draw [->-] (G.east |- Y) to (Y); \draw [->-] (R) to (G.east |- R);
\end{tikzpicture} \end{center}

We do not have space to define the entire categorical structure of $\mathbf{OG}$, instead referring the reader to \cite{hedges_etal_compositional_game_theory,hedges_towards_compositional_game_theory}, but we define categorical composition as an illustration.
Let $\mathcal G : \binom X S \to \binom Y R$ and $\mathcal H : \binom Y R \to \binom Z Q$ be open games.
The composite open game $\mathcal H \circ \mathcal G : \binom X S \to \binom Z Q$ is defined as follows:
\begin{itemize}
	\item The set of strategy profiles is $\Sigma_{\mathcal H \circ \mathcal G} = \Sigma_\mathcal G \times \Sigma_\mathcal H$
	\item The play function is $\mathbf P_{\mathcal H \circ \mathcal G} ((\sigma, \tau), x) = \mathbf P_\mathcal H (\tau, \mathbf P_\mathcal G (\sigma, x))$
	\item The coplay function is $\mathbf C_{\mathcal H \circ \mathcal G} ((\sigma, \tau), x, q) = \mathbf C_\mathcal G (\sigma, x, \mathbf C_\mathcal H (\tau, \mathbf P_\mathcal G (\sigma, x), q))$
	\item The equilibrium function is
	\[ \mathbf E_{\mathcal H \circ \mathcal G} (h, k) = \{ (\sigma, \tau) \mid \sigma \in \mathbf E_\mathcal G (h, k_\tau) \text{ and } \tau \in \mathbf E_\mathcal H (\mathbf P_\mathcal G (\sigma, h), k) \} \]
	where $k_\tau : Y \to R$ is given by $k_\tau (y) = \mathbf C_\mathcal H (\tau, y, k (\mathbf P_\mathcal H (\tau, y)))$.
\end{itemize}

A \emph{zero-player open game} is an open game $\mathcal G$ is one which has exactly one strategy profile, which is an equilibrium for every context.

$\mathbf{OG}$ is not a compact closed category, but it does have a `partial duality' which can be defined on objects by $\binom X S^* = \binom S X$.
For every $X$ there is a canonical (zero-player) open game $\varepsilon_X : \binom X X \to I$, but there is no canonical dual $I \to \binom X X$.
We denote $\varepsilon_X$ by the string diagram
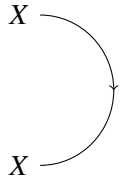
\begin{center} \begin{tikzpicture}
	\node (X1) at (0, 2) {$X$}; \node (X2) at (0, 0) {$X$};
	\draw [->-] (X1) to [out=0, in=90] (1.25, 1) to [out=-90, in=0] (X2);
\end{tikzpicture} \end{center}
Given a set $X$, we write $\overline X$ for $\binom X 1$ and $\underline X$ for $\binom 1 X$.
A function $f : X \to Y$ has both covariant and contravariant liftings as zero-player open games, which we call $\overline f : \overline X \to \overline Y$ and $\underline f : \underline Y \to \underline X$, which are dual in this sense; but no other open game has a dual.
The former has play function $\mathbf P_{\overline f} (*, x) = f (x)$, and the latter has coplay function $\mathbf C_{\underline f} (*, *, x) = f (x)$.
We denote these by
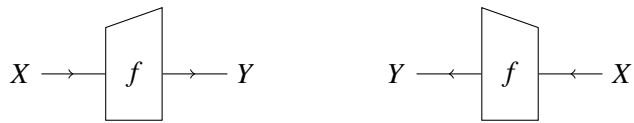
\begin{center} \begin{tikzpicture}
	\node (X1) at (0, 0) {$X$}; \node (Y1) at (3, 0) {$Y$};
	\node [trapezium, trapezium left angle=0, trapezium right angle=75, shape border rotate=90, trapezium stretches=true, minimum height=.75cm, minimum width=1.5cm, draw] (f1) at (1.5, 0) {$f$};
	\draw [->-] (X1) to (f1); \draw [->-] (f1) to (Y1);
	\node (X2) at (8, 0) {$X$}; \node (Y2) at (5, 0) {$Y$};
	\node [trapezium, trapezium left angle=75, trapezium right angle=0, shape border rotate=270, trapezium stretches=true, minimum height=.75cm, minimum width=1.5cm, draw] (f2) at (6.5, 0) {$f$};
	\draw [->-] (X2) to (f2); \draw [->-] (f2) to (Y2);
\end{tikzpicture} \end{center}

\subsection{Open games by example}

The easiest way to show the expressive power of open games is by examples of closed games $\mathcal G : I \to I$, which are represented by string diagrams with no open strings passing the left or right boundaries.

Given agents $\mathcal A_i : I \to \binom{X_i}{\mathbb R}$ as defined above (where $\mathcal A_i$ is a decision of an agent making an observation from $1 = \{ * \}$ and a choice from $X_i$), the $n$-fold monoidal product $\displaystyle \bigotimes_{i = 1}^n \mathcal A_i : I \to \begin{pmatrix} \prod_{i = 1}^n X_i \\ \mathbb R^n \end{pmatrix}$ represents a game in which $n$ player make choices simultaneously.
The set of strategy profiles of this game is $\Sigma_{\bigotimes_{i = 1}^n \mathcal A_i} = \prod_{i = 1}^n X_i$ and, given a payoff function $k : \prod_{i = 1}^n X_i \to \mathbb R^n$, the equilibrium set $\mathbf E_{\bigotimes_{i = 1}^n \mathcal A_i} (*, k)$ is the set of Nash equilibria of the $n$-player game with payoff function $k$, namely
\[ \mathbf E_{\bigotimes_{i = 1}^n \mathcal A_i} (*, k) = \left\{ \left. \sigma : \prod_{i = 1}^n X_i\ \right|\ k (\sigma)_i \geq k (\sigma [i \mapsto x_i])_i \text{ for all } 1 \leq i \leq n \text{ and } x_i : X_i \right\} \]

We can extend this to a closed game by fixing some particular $k$, by lifting $k$ as a function and combining it with $\varepsilon$.
For example, suppose we restrict to $n = 2$ and pick some payoff function $k : X \times Y \to \mathbb R^2$.
This is the class of `bimatrix games' and contains many famous examples such as the prisoner's dilemma and chicken, depending on the choice of $X$, $Y$ and $k$.

The 2-player closed game with payoff function $k$ is represented by the string diagram
	
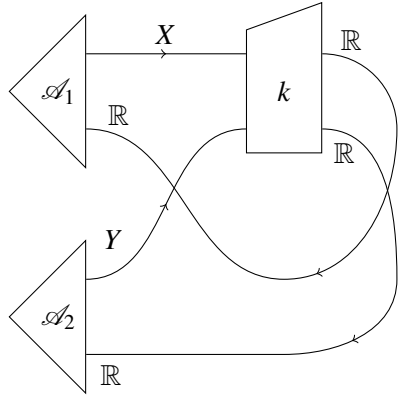
\begin{center} \begin{tikzpicture}
		\node [isosceles triangle, isosceles triangle apex angle=90, shape border rotate=180, minimum width=2cm, draw] (D1) at (0, 3) {$\mathcal A_1$};
		\node [isosceles triangle, isosceles triangle apex angle=90, shape border rotate=180, minimum width=2cm, draw] (D2) at (0, 0) {$\mathcal A_2$};
		\node [trapezium, trapezium left angle=0, trapezium right angle=75, shape border rotate=90, trapezium stretches=true, minimum height=1cm, minimum width=2cm, draw] (U) at (3, 3) {$k$};
		\node (d1) at (0, -.5) {}; \node (d2) at (0, .5) {}; \node (d3) at (0, 2.5) {}; \node (d4) at (0, 3.5) {}; \node (d5) at (0, 1) {}; \node (d6) at (0, 2) {};
		\draw [->-] (D1.east |- d4) to node [above] {$X$} (U.west |- d4);
		\draw [->-] (D2.east |- d2) to [out=0, in=180] node [above=5pt, very near start] {$Y$} (U.west |- d3);
		\draw [->-] (U.east |- d4) to [out=0, in=90] node [above, near start] {$\mathbb R$} (4.5, 2.5) to [out=-90, in=0] (3, .5) to [out=180, in=0] node [above, very near end] {$\mathbb R$} (D1.east |- d3);
		\draw [->-] (U.east |- d3) to [out=0, in=90] node [below, very near start] {$\mathbb R$} (4.5, .5) to [out=-90, in=0] (3, -.5) to node [below, very near end] {$\mathbb R$} (D2.east |- d1);
	\end{tikzpicture} \end{center}
Recall that a closed game is defined by a set of strategy profiles and a subset of Nash equilibria.
The closed game $\mathcal G : I \to I$ defined by this string diagram has strategy profiles $\Sigma_\mathcal G = X \times Y$, and Nash equilibria
\[ \mathbf E_\mathcal G = \{ (x, y) : X \times Y \mid k (x, y)_1 \geq k (x', y)_1 \text{ for all } x' : X \text{, and } k (x, y)_2 \geq k (x, y')_2 \text{ for all } y' : Y \} \]

Next, given a decision $\mathcal A : \binom X 1 \to \binom{Y}{\mathbb R}$ that observes an $X$ and chooses a $Y$, we form the composite $\mathcal A^\Delta : \binom X 1 \to \binom{X \times Y}{\mathbb R}$ defined by the string diagram
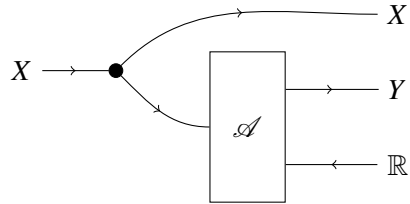
\begin{center} \begin{tikzpicture}
	\node (X) at (-3, .75) {$X$}; \node (Y) at (2, .5) {$Y$}; \node (R) at (2, -.5) {$\mathbb R$}; \node (X2) at (2, 1.5) {$X$};
	\node [rectangle, minimum height=2cm, minimum width=1cm, draw] (G) at (0, 0) {$\mathcal A$};
	\node [circle, scale=.5, fill=black, draw] (m) at (-1.75, .75) {};
	\draw [->-] (X) to (m); \draw [->-] (m) to [out=-45, in=180] (G); \draw [->-] (m) to [out=45, in=180] (X2);
	\draw [->-] (G.east |- Y) to (Y); \draw [->-] (R) to (G.east |- R);
\end{tikzpicture} \end{center}
where the node denotes the lifting of the copying function $\Delta : X \to X \times X$.
This important gadget has set of strategy profiles $\Sigma_{\mathcal A^\Delta} = X \to Y$, play function $\mathbf P_{\mathcal A^\Delta} (\sigma, x) = (x, \sigma (x))$ and equilibrium function
\[ \mathbf E_{\mathcal A^\Delta} (h, k) = \{ \sigma : X \to Y \mid \sigma (h) \in \arg\max k (h, -) \} \]

Using this gadget we can build sequential games.
For example, a 2-player sequential game in which the first player chooses from $X$, the second player observes this perfectly before choosing a $Y$, and payoffs being given by $k : X \times Y \to \mathbb R^2$ can be built from the decision $\mathcal A_1 : I \to \binom{X}{\mathbb R}$ and $\mathcal A_2 : \binom{X}{1} \to \binom{Y}{\mathbb R}$ using the string diagram
	
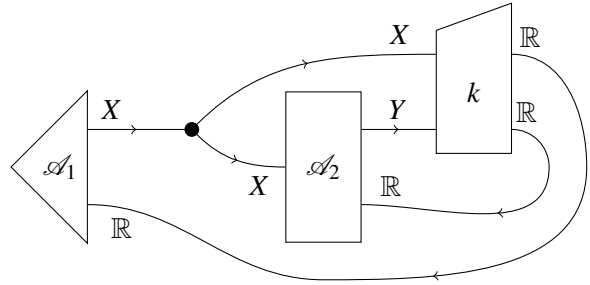
\begin{center} \begin{tikzpicture}
		\node [isosceles triangle, isosceles triangle apex angle=90, shape border rotate=180, minimum width=2cm, draw] (D1) at (0, 0) {$\mathcal A_1$};
		\node [circle, scale=.5, fill=black, draw] (m) at (1.75, .5) {};
		\node [rectangle, minimum height=2cm, minimum width=1cm, draw] (D2) at (3.5, 0) {$\mathcal A_2$};
		\node [trapezium, trapezium left angle=0, trapezium right angle=75, shape border rotate=90, trapezium stretches=true, minimum height=1cm, minimum width=2cm, draw] (q) at (5.5, 1) {$k$};
		\node (d1) at (0, -.5) {}; \node (d2) at (0, 1.5) {};
		\draw [->-] (D1.east |- m) to node [above, near start] {$X$} (m);
		\draw [->-] (m) to [out=45, in=180] node [above, very near end] {$X$} (q.west |- d2);
		\draw [->-] (m) to [out=-45, in=180] node [below, near end] {$X$} (D2);
		\draw [->-] (D2.east |- m) to node [above] {$Y$} (q.west |-m);
		\draw [->-] (q.east |- d2) to [out=0, in=90] node [above, very near start] {$\mathbb R$} (7, 0) to [out=-90, in=0] (3.5, -1.5) to [out=180, in=0] node [below, very near end] {$\mathbb R$} (D1.east |- d1);
		\draw [->-] (q.east |- m) to [out=0, in=90] node [above, near start] {$\mathbb R$} (6.5, 0) to [out=-90, in=0] node [above, very near end] {$\mathbb R$} (D2.east |- d1);
	\end{tikzpicture} \end{center}
	
Similarly, if the second player cannot perfectly observe the first move $x$ but only its image under some function $f : X \to Z$, we can model this using the string diagram obtained from the previous one by inserting an $f$-labelled node between the copying node and $\mathcal A_2 : \binom{Z}{1} \to \binom{Y}{\mathbb R}$.
%
As a typical example we impose an equivalence relation $\sim$ on $X$, and let $f : X \to X / \sim$ be the projection onto the quotient.
Such equivalence classes are known as `information sets' in game theory.
If $f : X \to 1$ then we recover a simultaneous game, since deleting is the counit for the black comonoid.

Note that each of the previous examples readily generalises to games of $n$ players.

\section{Process grammar}\label{sec:process-grammar}

We analyse the methodology of categorical compositional distributional semantics, arguing that we need to modify the concept of a pregroup grammar in order to compensate for the \emph{semantic} category lacking a compact closed structure.
Specifically, we argue that pregroup grammars rely on the semantic category satisfying state-process duality, a feature not present in the category of open games.
While the grammatical parsing of sentences using a pregroup grammar relies only on the \emph{counits} $x^l x \leq 1$, $x x^r \leq 1$ of a pregroup's compact closed structure \cite[corollary 1]{lambek1997}, state-process duality in the semantic category relies on the \emph{units} $I \to X \otimes X^*$.
In order to capture this Lambek defines \emph{protogroups} similarly to pregroups, but having only counits.

Consider the typing of a noun phrase consisting of an adjective and a noun, such as \emph{large slabs}.
Typically, we have a primitive type of noun phrases $n \in T$, and the dictionary $\mathcal D$ contains the entries $(\text{slabs}, n)$ and $(\text{large}, n n^l)$.
We have a semantic functor $\semantics - : \mathcal C_T \to \mathbf{FVect}$, whose definition includes a choice of vector space $N = \semantics n$, the \emph{noun space}.
By functorality, the type $n n^l$ of \emph{large} is sent to the tensor product $N \otimes N^*$.
However, we might imagine that the the semantics of an adjective is fundamentally a (linear) \emph{function}, where for example the semantics of \emph{large} is a function that takes the semantics of a noun phrase $x$ to the semantics of the noun phrase \emph{large} $x$.
The unit $\eta_N : I \to N \otimes N^*$ in $\mathbf{FVect}$ (where $I = \mathbb R$) is what allows us to encode a linear map $f : N \to N$ as a \emph{vector} $\overline f \in N \otimes N^*$, such that function application $f (v)$ can be recovered (after encoding vectors in $V$ as linear maps $I \to V$) via the \emph{counit} $\varepsilon_N : N^* \otimes N \to I$
\[ I \cong I \otimes I \xrightarrow{\overline f \otimes v} (N \otimes N^*) \otimes N \cong N \otimes (N^* \otimes N) \xrightarrow{N \otimes \varepsilon_N} N \otimes I \cong N \]

This entire methodology fails when we replace $\mathbf{FVect}$ with a different semantic category, such as $\mathbf{OG}$, which does not have state-process duality.
($\mathbf{OG}$ is additionally not monoidal closed, which also rules out using a Lambek calculus.)
In this section we develop an alternative.

Notice that the following sets are equal: (1) Elements of the free pregroup on $T$; (2) Objects of the strict free compact closed category with generating objects $T$; (3) Elements of the free protogroup on $T$; (4) Objects of the free strict teleological category with generating objects $T$.
See \cite{lambek1997,preller_lambek_free_compact_2_categories,kelly80,hedges_coherence_lenses_open_games} for various characterisations of these free structures.
A `teleological category' is a category whose abstract structure is similar to $\mathbf{OG}$, in particular having counits but no units; posetal teleological categories are protogroups.

\begin{definition}
	A \emph{process grammar} is a triple $G = (\mathcal L, T, s, \mathcal D)$ where
	\begin{itemize}
		\item $\mathcal L$ is the \emph{lexicon}, i.e. the set of words
		\item $T$ is the set of primitive types
		\item $s \in P_T$ is the \emph{sentence type}, where $P_T$ is (the set of elements of) the free pregroup on $T$
		\item The dictionary $\mathcal D$ is a subset $\mathcal D \subseteq \mathcal L \times P_T \times P_T$
	\end{itemize}
\end{definition}

The difference between this definition and the standard definition of a pregroup grammar (see section \ref{sec:DisCoCat}) is that words are associated to \emph{pairs} of types, rather than single types.
Given a dictionary entry $(w, t) \in \mathcal D$ in a pregroup grammar, we view the semantics $\semantics{w}$ as a \emph{state} $I \to \semantics{t}$.
Given a dictionary entry $(w, t, t') \in \mathcal D$ in a process grammar, we view the semantics $\semantics{w}$ as a \emph{process} $\semantics{t} \to \semantics{t'}$. It was not necessary in the pregroup grammar to reprent the semantics $\semantics{w}$ as a process, since in a compact closed category a morphism $A \rightarrow B$ can always be represented as a morphism $I \rightarrow B \otimes A^*$, and hence we only need keep track of one object. In contrast, in a process grammar we need both objects $A$ and $B$.
We consider every pregroup grammar $(\mathcal L, T, s, \mathcal D)$ as a process grammar $(\mathcal L, T, s, \mathcal D')$ using the dictionary $\mathcal D' = \{ (w, 1, t) \mid (w, t) \in \mathcal D \}$.

\begin{definition}
	Let $G = (\mathcal L, T, s, \mathcal D)$ be a process grammar.
	We write $\mathcal C_G$ for the free teleological category with
	\begin{itemize}
		\item The set of generating objects is $T$
		\item The set of generating morphisms\footnote{Formally, a teleological category distinguishes `dualisable morphisms' and `non-dualisable morphisms'. For simplicity, we consider all generators of $\mathcal C_G$ to be non-dualisable.} is $\{ w_{t, t'} : t \to t' \mid (w, t, t') \in \mathcal D \}$
	\end{itemize}
\end{definition}

The free teleological category is characterised in \cite{hedges_coherence_lenses_open_games} as a category whose morphisms are equivalence classes of suitable string diagrams.

With a pregroup grammar $G = (\mathcal L, T, s, \mathcal D)$, parsing a sentence $w_1 \ldots w_n$ is a 2-stage process:
\begin{enumerate}
	\item Choose dictionary entries $(w_i, t_i) \in \mathcal D$
	\item Choose a morphism $t_1 \cdots t_n \to s$ in $\mathcal C_T$, where $\mathcal C_T$ is the free compact closed category with generating objects $T$
\end{enumerate}
With a process grammar $\mathcal G$, these two stages become conflated: a choice of morphism $1 \to s$ in $\mathcal C_G$ contains both a choice of type for each word, and a reduction to the sentence type, since the types of each word represent the processes that comprise the reduction.
This method of parsing is also considered in \cite{toumi_categorical_compositional_distributional_questions}, where its equivalence to the previously described method is proved.

There is a problem with this: the sentence has disappeared!
Out of all of the morphisms $1 \to s$ in $\mathcal C_G$, how do we recognise those which are parsings of a particular sentence $w_1 \ldots w_n$? We must identify the morphisms $1 \to s$ that corresponds to the sentence we actually want to parse.
A solution to this is to present the free teleological category $\mathcal C_G$ using a sequent calculus, which can be done in a standard way by identifying suitable proofs \cite{blute_scott_category_theory_linear_logicians,lambek_scott_introduction_higher_order_categorical_logic}, where the categorical composition is cutting proofs together.

Specifically, consider a noncommutative linear calculus whose judgements are $\varphi \vdash \psi$ for $\varphi, \psi \in P_T$.
For each $t \in T$ we have three identity axioms:
\[ (l\text{-id})\frac{}{t^l \vdash t^l} \qquad\qquad (\text{id})\frac{}{t \vdash t} \qquad\qquad (r\text{-id})\frac{}{t^r \vdash t^r} \]
For each $(w, \varphi, \psi) \in \mathcal D$ we have an axiom
\[ (w_{\varphi, \psi})\frac{}{\varphi \vdash \psi} \]
Only two of the four axioms for negation are present, reflecting that our grammar corresponds to protogroups rather than pregroups:
\[ (l\text{-intro})\frac{\varphi \vdash t \psi}{t^l \varphi \vdash \psi} \qquad\qquad (r\text{-intro})\frac{\varphi \vdash \psi t}{\varphi t^r \vdash \psi} \]
Finally, we have the cut and $\otimes$-introduction rules:
\[ (\text{cut})\frac{\varphi \vdash \psi \qquad \psi \vdash \chi}{\varphi \vdash \chi} \qquad\qquad (\otimes\text{-intro})\frac{\varphi \vdash \psi \qquad \varphi' \vdash \psi'}{\varphi \varphi' \vdash \psi \psi'} \]

With this setup, we can define a parsing of the sentence $w_1 \ldots w_n$ to be a derivation of $\vdash s$ such that, if we read the non-identity axioms used in the derivation from left to right, they are labelled by the $w_i$ in the correct order.
Derivations then witness morphisms in $\mathcal C_G$, and so we can define the sentence semantics by applying a semantic functor directly to the derivation.

As an example, suppose $\mathcal L$ contains the words \emph{bring large slabs}, $T$ contains $n$ and $s$, and $\mathcal D$ contains the entries $(\text{bring}, 1, s n^l)$, $(\text{large}, n^l, n^l)$ and $(\text{slabs}, 1, n)$. Here is an example of a derivation of the sentence \emph{bring large slabs}:
\begin{prooftree}
				\AxiomC{}\LeftLabel{(bring)}\UnaryInfC{$\vdash s n^l$}
			\AxiomC{}\LeftLabel{(id)}\UnaryInfC{$s \vdash s$}
	\AxiomC{}\LeftLabel{(large)}\UnaryInfC{$n^l \vdash n^l$}
	\AxiomC{}\LeftLabel{(slabs)}\UnaryInfC{$\vdash n$}
		\LeftLabel{($l$-intro)}\UnaryInfC{$n^l \vdash$}
			\LeftLabel{(cut)}\BinaryInfC{$n^l \vdash$}
				\LeftLabel{($\otimes$-intro)}\BinaryInfC{$s n^l \vdash s$}
					\LeftLabel{(cut)}\BinaryInfC{$\vdash s$}
\end{prooftree}

Note that the restriction to the free teleological category does not restrict the sentences which are grammatical: Lambek's switching lemma \cite{Lambek01} states that a sequence of epsilon and identity maps suffice for the representation
of the grammatical structure of any sentence in a pregroup grammar, and hence the grammatical structure of any sentence in a pregroup grammar may also be represented in the teleological grammar.

\section{Language-games as functors}

Given the pieces we have set up, with the power of category theory our model can be stated very simply: We study functors from a protogroup (or more precisely, a free teleological category) to the category of open games.
More precisely, we require that this functor preserves all of the structure of the protogroup, so it should be a monoidal functor and also respect duals.
Functors with the required properties were named \emph{teleological functors} in \cite{hedges_coherence_lenses_open_games}, although we again need to generalise to the non-symmetric case.

\begin{definition}
	Let $G = (\mathcal L, T, s, \mathcal D)$ be a process grammar.
	A \emph{functorial language-game} $\semantics -$ on $G$ is determined by the following data:
	\begin{itemize}
		\item For each type $t \in T$, a pair of sets $\semantics t$
		\item For each dictionary entry $(w, t_1, t_2) \in \mathcal D$, a choice of open game $\semantics{(w, t_1, t_2)} : \semantics{t_1} \to \semantics{t_2}$
	\end{itemize}
	Here $\semantics{t_i}$ is the object of $\mathbf{OG}$ (pair of sets) computed functorially in the evident way: $\semantics{t t'} = \semantics{t_1} \otimes \semantics{t_2}$ (componentwise cartesian product), and $\semantics{t^l} = \semantics{t^r} = \semantics{t}^*$ (swapping the pair).
\end{definition}
%

A functorial language-game $\semantics -$ on $G$ induces a functor $\semantics - : \mathcal C_G \to \OG$ in an evident way, since the data specifies its action on the generators of $\mathcal C_G$.
In particular we take $\semantics{\varepsilon_t} = \varepsilon_{\semantics t}$.

A functorial language-game associates a game to every sentence generated by a process grammar, in which the structure of the information flow in the game exactly reflects the grammatical structure of the sentence.
In practice this is a very strong restriction, and it is unclear whether it is actually desirable; in this paper we are demonstrating that it is in principle possible, at least in simple cases.

In particular, consider a sentence of the form $w_1 \ldots w_i \ldots w_j \ldots w_n$, in particular with $w_i$ appearing before $w_j$ in the list of words.
In the resulting open game, the part corresponding to $w_i$ happens temporally before the part corresponding to $w_j$.
Depending on the types, players in $w_j$ might or might not be able to observe the result of $w_i$; but there is no possible way for players in $w_i$ to observe $w_j$.
This is awkward, and is also very sensitive to word order in the language we are modelling (English in our case).

For example, in the next section we will model sentences such as \emph{bring slabs}, where \emph{bring} contains an agent who utters the order with a strategic preference for slabs to be brought.
However, the agent representing \emph{bring} is unable to observe \emph{slabs}!
Instead, we will define \emph{slabs} as a zero-player open game, which means that the agent in \emph{bring} is able to perfectly anticipate it.
This is a subtle distinction, related to the strange categorical structure of $\mathbf{OG}$, but one simple consequence is that the agent's strategy cannot be a function that chooses an order for every possible noun phrase.

\section{Example}\label{sec:example}

We will build an example based on \cite{wittgenstein_philosophical_investigations}'s example of the master builder and apprentice.
Consider a universe that contains \emph{slabs} and \emph{planks}, both in two sizes, \emph{small} and \emph{large}.
The master can order the apprentice either to \emph{bring} an object, or to \emph{cut} it.

We build a simple categorial grammar to describe the master's commands.
(Imperative) sentences in this grammar will be interpreted as open games containing a single player, representing the master, who makes a move representing the order and receives utility if the appropriate action is carried out.
This open game can be \emph{closed} by composing it with a second player, the apprentice.
We can study the Nash equilibria of the resulting game.

We define a process grammar $G = (\mathcal L, T, s, \mathcal D)$ as follows.
Let $\mathcal L = \{ \text{slabs}, \text{planks}, \text{large}, \text{bring}, \text{cut} \}$.
Let $T = \{ n, s \}$ where $n$ is the basic type of noun phrases and $s$ is the basic type of imperative sentences.
The dictionary is given by
\[ \text{slabs} : 1 \leq n \qquad\qquad \text{planks} : 1 \leq n \qquad\qquad \text{large} : n^l \leq n^l \]
\vspace{-.75cm}
\[ \text{bring} : 1 \leq s n^l \qquad\qquad \text{cut} : 1 \leq s n^l \]
where $w : x \leq y$ means $(w, x, y) \in \mathcal D$.

We interpret the nouns and adjectives as zero-player open games lifted from a simple Montague-like semantics.
Let $U$ be a semantic universe, with chosen subsets $S, P, L \subseteq U$ of objects that are slabs, planks, and large things. We begin building a semantic functor $\semantics - : \mathcal C_G \to \mathbf{OG}$.

We firstly interpret $\semantics n = \overline{\mathcal P (U)}$.
The nouns \emph{slabs} and \emph{planks} are interpreted as the zero-player open games containing the constant function returning the subsets of $S$ and $P$ respectively.
So $\semantics{\text{slabs}} : I \to \overline{\mathcal P (U)}$ is the zero-player open game with play function $\mathbf P_{\semantics{\text{slabs}}} (*, *) = S$, and similarly for \emph{planks}.

Consider the function $\cap L : \mathcal P (U) \to \mathcal P (U)$ given by $X \mapsto X \cap L$, i.e. it takes a set of things to the subset of those things which are large.
\emph{Large} has the grammatical process-type $n^l \leq n^l$, so $\semantics{\text{large}}$ must be an open game $\underline{\mathcal P (U)} \to \underline{\mathcal P (U)}$.
We define $\semantics{\text{large}} = \underline{\cap L}$.

We model an imperative verb as a player (representing the master) choosing from a choice of actions, which we think of as utterances in some language.
This language could be defined to be the same language defined by the categorial grammar we are considering, but in general it need not be.
Let $O$ be a set of possible utterances (orders) in this language, which we assume nothing about; defining $O$ to be the set of well-typed sentences of $G$ is one possible example.

We need to define a preference structure for the master.
He receives two outcomes: the first, which will be a constant, represents the set of objects referred to by the subsequent noun phrase.
The second represents the action done by the apprentice.
If the verb in question is \emph{bring} then the outcome $(S, a)$ is preferred, where $S \subseteq U$ and $a$ is an action, precisely when $a$ represents bringing some element of $S$.
Similarly if the verb is \emph{cut} then $(S, a)$ is preferred when $a$ represents cutting some element of $S$.

Let $A = \{ B (x) \mid x \in U \} \cup \{ C (x) \mid x \in U \}$ be the set of possible actions, where $B (x)$ represents bringing $x$ and $C (x)$ represents cutting it.
We define the sentence space to be $\semantics s = \binom O A$.
This means that a sentence with derivation $1 \to s$ is represented as an open game $I \to \binom O A$ which chooses an element of $O$ with outcomes in $A$.

The verb \emph{bring} has grammatical process type $1 \leq s n^l$, which means its semantics must be an open game with type $\semantics{\text{bring}} : I \to \binom O A \otimes \underline{\mathcal P (U)} = \binom{O}{A \times \mathcal P (U)}$.
We define it as the following open game:
\begin{itemize}
	\item The set of strategy profiles is $\Sigma_{\semantics{\text{bring}}} = O$
	\item The play function $\mathbf P_{\semantics{\text{bring}}} : \Sigma_{\semantics{\text{bring}}} \times 1 \to O$ is given by $\mathbf P_{\semantics{\text{bring}}} (o, *) = o$
	\item The coplay function $\mathbf C_{\semantics{\text{bring}}} : \Sigma_{\semantics{\text{bring}}} \times 1 \times A \times \mathcal P (U) \to 1$ is given by $\mathbf C_{\semantics{\text{bring}}} (o, *, a, S) = *$
	\item The equilibrium function $\mathbf E_{\semantics{\text{bring}}} : 1 \times (O \to A \times \mathcal P (U)) \to \mathcal P (\Sigma_{\semantics{\text{bring}}})$ is given by
	\[ \mathbf E_{\semantics{\text{bring}}} (*, k) = \{ o \mid k (o)_1 = B (x) \text{ for some } x \in k (o)_2 \} \]
\end{itemize}
The open game $\semantics{\text{cut}} : I \to \binom{O}{A \times \mathcal P (U)}$ is defined in the same way, except that $B (x)$ is changed to $C (x)$.

Consider the sentence \emph{bring large slabs}, with the derivation $1 \leq s$ given at the end of section \ref{sec:process-grammar}.
Applying the functor $\semantics -$ produces the open game $\semantics{\text{bring large slabs}} : I \to \binom O A$ represented by the diagram:
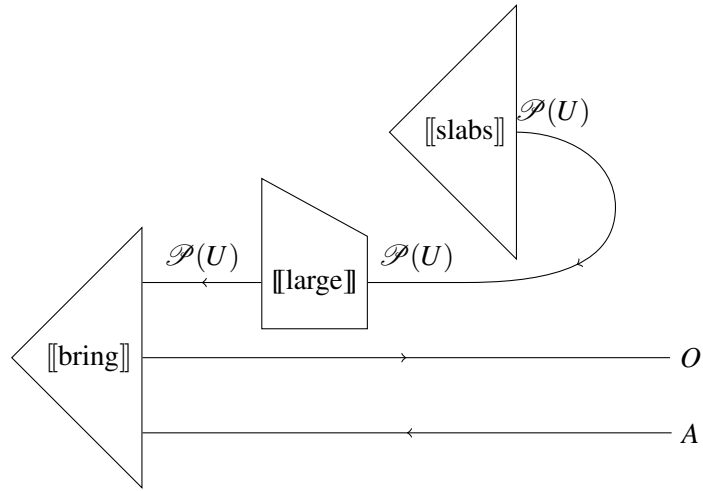
\begin{center} \begin{tikzpicture}
	\node (bring) [isosceles triangle, isosceles triangle apex angle=90, shape border rotate=180, minimum width=2cm, draw] at (-2, 0) {$\semantics{\text{bring}}$};
	\node (large) [trapezium, trapezium left angle=75, trapezium right angle=0, shape border rotate=270, trapezium stretches=true, minimum height=1cm, minimum width=2cm, draw] at (1, 1) {\semantics{\text{large}}};
	\node (slabs) [isosceles triangle, isosceles triangle apex angle=90, shape border rotate=180, minimum width=2cm, draw] at (3, 3) {$\semantics{\text{slabs}}$};
	\node (O) at (6, 0) {$O$}; \node (A) at (6, -1) {$A$};
	\draw [->-] (slabs) to [out=0, in=90] node [above, near start] {$\mathcal P (U)$} (5, 2) to [out=-90, in=0] (3, 1) to node [above] {$\mathcal P (U)$} (large);
	\draw [->-] (large) to node [above] {$\mathcal P (U)$} (bring.east |- large);
	\draw [->-] (bring) to (O); \draw [->-] (A) to (bring.east |- A);
\end{tikzpicture} \end{center}
Explicitly, this open game is given by the following data, which can be computed compositionally given the operations on $\mathbf{OG}$:
\begin{itemize}
	\item The set of strategy profiles is $\Sigma_{\semantics{\text{bring large slabs}}} = O$
	\item The play function is $\mathbf P_{\semantics{\text{bring large slabs}}} (o, *) = o$
	\item The equilibrium function, which takes a parameter $k : O \to A$, is
	\[ \mathbf E_{\semantics{\text{bring large slabs}}} (k) = \{ o : O \mid k (o) = B (x) \text{ for some } x \in S \cap L \} \]
\end{itemize}
In other words, given a continuation $k : O \to A$ that converts orders into actions (which abstracts over the behaviour of an as-yet-unspecified apprentice), the good orders from the master's perspective are the ones that the continuation will convert into the action of bringing some slab.
It should be noted that this open game contains no built-in relationship between the orders $O$ and actions $A$, but rather takes this as a parameter through $k$; that is to say, our model of the master has not fixed a semantics of orders.

This open game defines the master's half of the situation, abstracting out everything else into the continuation $k$.
In order to produce a genuine game we must compose with another open game representing the decision made by the apprentice.
In our setup this is an extra post-processing step, and has not been produced functorially from the sentence's grammar.

In order to make the apprentice into a game-theoretic agent, we must consider what his objectives are.
In Wittgenstein's example the master utters the words ``Bring large slabs'', and then the apprentice brings slabs; but even assuming the agent understands the master's language, why should he follow the order?
In order to model the situation game-theoretically, we must define the apprentice's objectives such that following the order becomes rational.

For now we will take a simple option: The apprentice has a built-in translation from $O$ to $A$, and has a preference to carry out the received order according to this endogenous language, in line with the Wittgenstein quote in section \ref{sec:language-games}.

%


Fix a function $f : O \times A \to \mathbb R$, which is the apprentice's judgement of the similarity between orders and outcomes.
We define the apprentice as an agent who, on receiving the order $o : O$, will choose some $a : A$ in order to maximise the similarity $f (o, a)$.
Let $\mathcal G : \binom O A \to I$ be the open game defined by the diagram
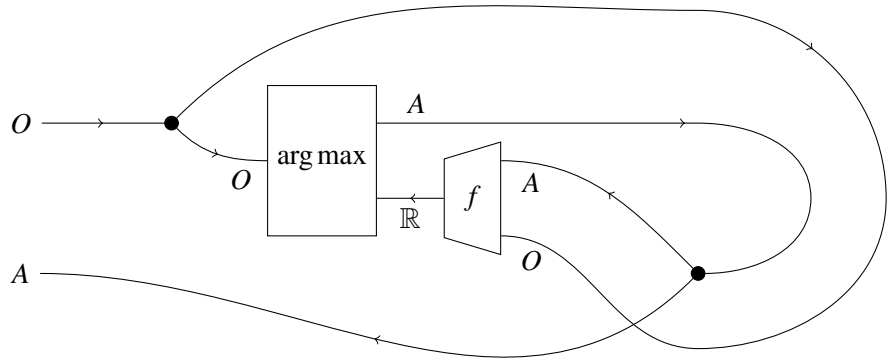
\begin{center} \begin{tikzpicture}
	\node (O) at (0, 2) {$O$}; \node (A) at (0, 0) {$A$};
	\node (copy) [circle, scale=.5, fill=black, draw] at (2, 2) {};
	\node (copy2) [circle, scale=.5, fill=black, draw] at (9, 0) {};
	\node (argmax) [rectangle, minimum height=2cm, draw] at (4, 1.5) {$\arg\max$};
	\node (f) [trapezium, trapezium left angle=75, trapezium right angle=75, shape border rotate=90, trapezium stretches=true, minimum height=.75cm, minimum width=1.5cm, draw] at (6, 1) {$f$};
	\node (dummy) at (0, .5) {};
	\draw [->-] (O) to (copy); \draw [->-] (copy) to [out=-45, in=180] node [below, near end] {$O$} (argmax);
	\draw [->-] (argmax.east |- O) to node [above, very near start] {$A$} (9, 2) to [out=0, in=90] (10.5, 1) to [out=-90, in=0] (copy2);
	\draw [->-] (copy2) to [out=135, in=0] node [below, very near end] {$A$} (f.east |- argmax); \draw [->-] (f) to node [below] {$\mathbb R$} (argmax.east |- f);
	\draw [->-] (copy2) to [out=-135, in=0] (A);
	\draw [->-] (copy) to [out=45, in=180] (9, 3.5) to [out=0, in=90] (11.5, 1) to [out=-90, in=0] (9, -1) to [out=180, in=0] node [below, very near end] {$O$} (f.east |- dummy);
\end{tikzpicture} \end{center}
\vspace{-20pt}
Concretely, it is given by the following data:
\begin{itemize}
	\item The set of strategy profiles is $\Sigma_\mathcal G = O \to A$
	\item The coplay function $\mathbf C_\mathcal G : \Sigma_\mathcal G \times O \times 1 \to A$ is $\mathbf C_\mathcal G (\sigma, o, *) = \sigma (o)$
	\item The equilibrium function $\mathbf E_\mathcal G : O \to \mathcal P (\Sigma_\mathcal G)$ is
	\[ \mathbf E_\mathcal G (o) = \{ \sigma \mid \sigma (o) \in \arg\max_{a : A}  f (o, a) \} \]
\end{itemize}

The final step is to form the composition $\mathcal G \circ \semantics{\text{bring large slabs}} : I \to I$, a closed game.
In a Nash equilibrium of the composite, the master uses the apprentice's built-in language in order to choose an order.
Concretely, the data specifying this game is:
\begin{itemize}
	\item The set of strategy profiles is $\Sigma_{\mathcal G \circ \semantics{\text{bring large slabs}}} = O \times (O \to A)$
	\item The Nash equilibria, which are now simply a subset of $\Sigma_{\mathcal G \circ \semantics{\text{bring large slabs}}}$, are
	\[ \{ (o, \sigma) \mid \sigma (o) = B (x) \text{ for some } x \in S \cap L \text{ and } \sigma (o) \in \arg\max_{a : A} f (o, a) \} \]
\end{itemize}

Since this is a sequential game we can compute equilibria by a method known as backward induction, which also reveals the way that the agents themselves might reason.
Although $o \mapsto \arg\max_{a : A} f (o, a)$ is a multi-valued function, each possible choice defines an optimal strategy $\sigma : O \to A$ for the apprentice.
These $\sigma$ are the ways that the master can deduce the apprentice might behave, given the apprentice's understanding of the language.
It may also be reasonable to assume that $f$ specifies the language unambiguously in the sense that each $\arg\max_{a : A} f (o, a)$ is unique, in which case $\sigma$ is uniquely defined.

A choice of $\sigma$ completely describes the apprentice's expected behaviour, so now we can reason as the master.
He would like to choose some order $o \in O$ such that $\sigma (o)$ is one of the actions that he will be satisfied with.
In this case, the set of good actions is $B [S \cap L]$ (the forward image of $S \cap L$ under $B$), so the set of good orders given $\sigma$ is $\sigma^{-1} [B [S \cap L]]$.

\section{Outlook}
The theory as developed so far describes a simple signalling game, but allows for compositional aspects of language to be included. 
Future work will introduce consequences for the apprentice, thereby enabling the master and apprentice to coordinate on a choice of language. 
There are several possibilities here.
A nice idea is that there are \emph{economic} consequences for the apprentice, such as being fired or having his pay docked.
A more linguistic angle is that the master verbally thanks or scolds the apprentice depending on the action, and the apprentice has preferences over these utterances.
Both of these have the attractive feature that the master and apprentice do not automatically share a language (precisely, a relationship between orders $O$ and actions $A$), but in each Nash equilibrium they successfully coordinate on a choice of language. Both of these require extending the game with further stages in the future, where the master moves again after the apprentice, possibly in an infinitely repeated game \cite{ghani_kupke_lambert_forsberg_compositional_treatment_iterated_open_games}.
Doing this functorially will require more extensive changes to the grammar, so we leave it for future work.
 Game theoretic approaches to linguistics show that convex concepts emerge from agents' interactions: using a more structured meaning space would allow us to investigate this.

\bibliographystyle{eptcs}
\bibliography{lg}

\end{document}